\documentclass[a4paper]{jpconf}
\usepackage{graphicx}
\begin{document}
\title{Brief review of double beta decay experiments}

\author{A S Barabash}
\address{National Research Center "Kurchatov Institute", Institute of Theoretical and Experimental Physics, B. Cheremushkinskaya 25, 117218 Moscow, Russia}

\ead{barabash@itep.ru\\}

\begin{abstract}
Best present experimental achievements in double beta decay are presented. Possible progress in this field in the near and far future is discussed.
\end{abstract}

\section{Introduction}
The detection and study of $0\nu\beta\beta$ 
decay may clarify the following problems of neutrino physics 
(see recent review \cite{BIL15}, for example): 
(i) lepton number non-conservation, (ii) neutrino nature: whether 
the neutrino is a Dirac or a Majorana particle, (iii) absolute 
neutrino mass scale, (iv) the 
type of neutrino mass hierarchy (normal, inverted, or quasidegenerate), 
(v) CP violation in the lepton sector (measurement of the Majorana 
CP-violating phases). 

Usually, three main modes of the 2$\beta$ decay are considered:

\begin{equation}
(A,Z) \rightarrow (A,Z+2) + 2e^{-} + 2\bar { \nu},
\end{equation}

\begin{equation}
(A,Z) \rightarrow (A,Z+2) + 2e^{-},
\end{equation}

\begin{equation}
(A,Z) \rightarrow (A,Z+2) + 2e^{-} + \chi^{0}(+ \chi^{0}),
\end{equation}

where A is the atomic number, Z is the charge of the nucleus, $e^-$ is the electron, 
$\bar { \nu}$ is an antineutrino, and $\chi^{0}$ is the Majoron.

Best present achievements for mentioned above processes are presented in tables 1-3.

\begin{table}[ht]
\label{Table1}
\caption{Average and recommended $T_{1/2}(2\nu)$ values (from
\cite{BAR15a}).}
\vspace{0.5cm}
%\rule[-2mm]{0mm}{5mm}
\begin{center}
\begin{tabular}{cc}
\hline
Isotope & $T_{1/2}(2\nu)$, yr \\
\hline
$^{48}$Ca & $4.4^{+0.6}_{-0.5}\cdot10^{19}$ \\
$^{76}$Ge & $1.65^{+0.14}_{-0.12} \cdot10^{21}$ \\
$^{82}$Se & $(0.92 \pm 0.07)\cdot10^{20}$ \\
$^{96}$Zr & $(2.3 \pm 0.2)\cdot10^{19}$ \\
$^{100}$Mo & $(7.1 \pm 0.4)\cdot10^{18}$ \\
$^{100}$Mo-$^{100}$Ru$(0^{+}_{1})$ & $6.7^{+0.5}_{-0.4}\cdot10^{20}$ \\
$^{116}$Cd & $(2.87 \pm 0.13)\cdot10^{19}$\\
$^{128}$Te & $(2.0 \pm 0.3)\cdot10^{24}$ \\
$^{130}$Te & $(6.9 \pm 1.3)\cdot10^{20}$ \\
$^{136}$Xe & $(2.19 \pm 0.06)\cdot10^{21}$ \\
$^{150}$Nd & $(8.2 \pm 0.9)\cdot10^{18}$ \\
$^{150}$Nd-$^{150}$Sm$(0^{+}_{1})$ & $1.2^{+0.3}_{-0.2}\cdot10^{20}$\\
$^{238}$U & $(2.0 \pm 0.6)\cdot10^{21}$  \\
$^{130}$Ba; ECEC(2$\nu$) & $\sim 10^{21}$  \\
\hline
\end{tabular}
\end{center}
\end{table}

\begin{table}[ht]
\label{Table2}
\caption{Best present results on $0\nu\beta\beta$ decay (limits at
90\% C.L.). To calculate $\langle m_{\nu} \rangle$ 
the NME from \cite{SUH15,SIM13,BAR15,RAT10,ROD10,MEN09,HOR15,MUS13,SON17}, 
phase-space factors from \cite{KOT12,STO15} and $g_A$ = 1.27 have been used.
In case of $^{150}$Nd NME from \cite{TER15,FAN15} 
and in case of $^{48}$Ca from \cite{IWA16} were used in addition.}
\vspace{0.5cm}
%\rule[-2mm]{0mm}{5mm}
\begin{center}
\begin{tabular}{ccccc}
\hline
Isotope & Q$_{2\beta}$, keV & $T_{1/2}$, y & $\langle m_{\nu} \rangle$, eV 
 & Experiment \\

\hline
$^{48}$Ca & 4267.98  & $>5.8\cdot10^{22}$ & $<3.1-15.4$ & CANDLES \cite{UME08} \\
$^{76}$Ge & 2039.00 & ${\bf >3.5\cdot10^{25}}$ & ${\bf <0.18-0.48}$ & GERDA-I+GERDA-II \cite{AGO16} \\
& & ($>5.2\cdot10^{25}$) & ($<0.15-0.39$) & \\
$^{82}$Se &2997.9 & $>3.6\cdot10^{23}$ & $<1-2.4$ & NEMO-
3 \cite{BAR11} \\
$^{96}$Zr & 3355.85 & $>9.2\cdot10^{21}$ & $<3.6-10.4$ & NEMO-3 \cite{ARN10} \\
$^{100}$Mo &3034.40 & $>1.1\cdot10^{24}$ & $<0.33-0.62$ & NEMO-
3 \cite{ARN15} \\
$^{116}$Cd &2813.50 & $>1.9\cdot10^{23}$ & $<1-1.8$ & AURORA \cite{DAN16} \\
$^{128}$Te &866.6 & $>1.5\cdot10^{24}$ & $2.3-4.6$ & Geochem. exp. (see \cite{BAR15a})  \\
$^{130}$Te &2527.52 & $>4\cdot10^{24}$ & $<0.26-0.97$ & CUORICINO + CUORE0 \cite{ALD16} \\
$^{136}$Xe & 2457.83 & ${\bf >0.5\cdot10^{26}}$ & ${\bf <0.09-0.24}$ & KamLAND-Zen
\cite{GAN16} \\
 & & ($>1.07\cdot10^{26}$) & ($<0.06-0.16$) & \\
$^{150}$Nd & 3371.38 & $>2\cdot10^{22}$ & $<1.6-5.3$ & NEMO-3 \cite{ARN16} \\
\hline
\end{tabular}
\end{center}
\end{table}

\begin{table}[ht]
\label{Table3}
\caption{Best present limits on $0\nu\chi^{0}\beta\beta$ decay
(ordinary Majoron) at 90\% C.L. To calculate $\langle g_{ee} \rangle$ 
the NME from \cite{SUH15,SIM13,BAR15,RAT10,ROD10,MEN09,HOR15,MUS13,SON17}, 
phase-space factors from \cite{KOT15} and $g_A$ = 1.27 have been used. In case of $^{150}$Nd NME from \cite{TER15,FAN15} 
and in case of $^{48}$Ca from \cite{IWA16} were used in addition.}
\vspace{0.5cm}
%\rule[-2mm]{0mm}{5mm}
\begin{center}
\begin{tabular}{cccc}
\hline
Isotope & $T_{1/2}$, y & $\langle g_{ee} \rangle$, $\times 10^{-5}$
 & Experiment \\  
\hline
$^{48}$Ca & $>4.6\cdot10^{21}$  & $< 8.6-43.1 $ & NEMO-3 \cite{ARN16b} \\
$^{76}$Ge & $>4.2\cdot10^{23}$  & $< 2.4-6.3 $ & GERDA-I \cite{AGO15} \\
$^{82}$Se & $>1.5\cdot10^{22}$  & $< 5.0-12.2 $ & NEMO-3 \cite{ARN06}\\
$^{96}$Zr & $>1.9\cdot10^{21}$  & $7.4-21.3$  & NEMO-3 \cite{ARN10} \\
$^{100}$Mo & $>4.4\cdot10^{22}$ & $< 1.7-3.1 $ & NEMO-3 \cite{ARN15} \\
$^{116}$Cd & $>1.1\cdot10^{22}$  & $4.5-8.0$  & AURORA \cite{DAN16} \\
$^{128}$Te & $>1.5\cdot10^{24}$  & $6.3-12.3$ & Geochem. exp. (see \cite{BAR15a})  \\
$^{130}$Te & $>1.6\cdot10^{22}$  & $< 4.6-17.4 $ & NEMO-3 \cite{ARN11} \\
$^{136}$Xe & $> 2.6\cdot10^{24}$ & $< 0.45-1.2 $ & KamLAND-Zen \cite{GAN12} \\
$^{150}$Nd & $>3\cdot10^{21}$  & $< 3.7-11.9 $ & NEMO-3 \cite{ARN16} \\
%$^{128}$Te & $>2\cdot10^{24}$(geochem)\cite{MAN91} & $<(0.7-
%1.6)\cdot10^{-4}$ & $<(1.9-2.4)\cdot10^{-4}$ \\
\hline
\end{tabular}
\end{center}
\end{table}

\section{Best recent results}

In 2016 new (record) limits for existence of neutrinoless double beta decay in experiments with $^{136}$Xe (KamLAND-Zen) and $^{76}$Ge (GERDA-II) have been obtained.

\subsection{$^{136}$Xe (KamLAND-Zen) \cite{GAN16}}
KamLAND-Zen experiment phase-2 (383 kg of the enriched xenon) has been finished in May 2016. Limit of $9.2\cdot10^{25}$ yr (90\% C.L.) for 534.5 days of measurements has been established. Combining this limit with result of a phase-1 \cite{GAN13} authors obtained limit of $1.07\cdot10^{26}$ yr (90\% C.L.), which corresponds to limit $\langle m_{\nu} \rangle$ $< 0.06-0.16$ eV. It must be emphasized that sensitivity of the experiment is $5\cdot10^{25}$ yr ($\langle m_{\nu} \rangle$ $\sim 0.09-0.24$ eV), and more strong limit is obtained thanks to big "negative" fluctuation of a background. To be conservative I recommend to use exactly this last value as most reliable and correct limit by this moment. Nevertheless for today this is the best limit on $\langle m_{\nu} \rangle$ among all experiments. 
     Now there is a preparation of measurements with 750 kg of the enriched xenon and a new (pure) internal balloon and sensitivity of the experiment will be increased up to $\sim 2\cdot10^{26}$ yr.

\subsection{$^{76}$Ge (GERDA-II)}

At the Neutrino-2016 Conference GERDA Collaboration reported the first results of  GERDA-II experiment with 35.8 kg of HPGe detectors manufactured of the enriched germanium.  10.8 kg$\cdot$yr of statistics has been obtain during May 2015 -- December 2016. The main achievement is the reduction of background in the double beta decay region to $\sim 10^{-3}$ c/keV$\cdot$kg$\cdot$yr (an order of magnitude below the background level obtained in GERDA-I \cite{AGO13}). As a result, there was no recorded events in neutrinoless double beta decay region. Combining the results of GERDA-I and GERDA-II a new limit on $T_{1/2}$(0$\nu$) for $^{76}$Ge has been obtained. Using two different methods the following results were obtained with a 90\% confidence level:

1) $T_{1/2}$(0$\nu$) $> 5.2\cdot10^{25}$ yr (sensitivity is $4\cdot10^{25}$ yr) (frequentist test statistics).

2) $T_{1/2}$(0$\nu$) $> 3.5\cdot10^{25}$ yr (sensitivity is $3\cdot10^{25}$ yr) (Bayesian method).\\
It seems that the second method provides more reliable and correct result.
Set statistics in GERDA-II experiment continues. The planned experiment sensitivity for 3 years of measurements is $\sim 10^{26}$ years.

\section{Prospects for the future experiments} 
Seven of the most developed and promising experiments which can
be realized within the next $\sim$(3-10) years are presented in table 4. The
estimation of the sensitivity to $\langle m_{\nu} \rangle$ is done using NMEs from \cite{SUH15,SIM13,BAR15,RAT10,ROD10,MEN09,HOR15,MUS13,SON17} and phase-space factor values from \cite{KOT12,STO15}. Actually the experiments specified in Table 4 don't settle all variety of the offered experimental
approaches to 2$\beta$ decay search. There is a set of other propositions (see reviews \cite{VER12,ELL12,SCH13}, for example). The part of them already is at a stage of development of prototypes with a mass of the studied isotope $\sim$ 1-10 kg (NEXT \cite{LAI16}, CANDLESS \cite{IID16}, 
LUCIFER \cite{ART16}, AMORE \cite{PAR16}, LUMINEU/LUCINEU \cite{BAR14,ARM16}).

In this Section I will try to predict the Future. In Section 3.1  the results which (I hope) will be obtained in 2017-2019 are presented. And in Sec. 3.2. possible dates for the start of data taking for a few large scale experiments are given.

\begin{table}[h]
%\setcaptionmargin{0mm} \onelinecaptionsfalse
%\captionstyle{flushleft} 
\caption{Seven most developed and promising projects. 
Sensitivity at 90\% C.L. for three (1-st step of GERDA and MAJORANA, 
first step of SuperNEMO, CUORE-0 and KamLAND-Zen) 
five (EXO-200, SuperNEMO, SNO+ and CUORE) and ten (EXO, full-scale GERDA and MAJORANA) 
years of measurements is presented. M - mass of isotopes.}
%\vspace{0.5cm}
%\rule[-2mm]{0mm}{5mm}
%\begin{center}
\begin{tabular}{llllll}
\hline
Experiment & Isotope & M, kg & Sensitivity & Sensitivity & Status \\
& &  & $T_{1/2}$, yr & $\langle m_{\nu} \rangle$, meV &  \\
\hline
CUORE \cite{ART15}  & $^{130}$Te & 200 & $9.5\times10^{25}$ & 53--200 & in progress \\  
GERDA \cite{AGO16a} & $^{76}$Ge & 35 & $1\times10^{26}$ & 110--280 & current \\
& & 1000 & $6\times10^{27}$ & 14--37 & R\&D \\ 
MAJORANA  & $^{76}$Ge & 30 & $1\times10^{26}$ & 110--280 & current \\
\cite{ABG14} & & 1000 & $6\times10^{27}$ & 14-37 & R\&D \\ 
EXO \cite{POC15} & $^{136}$Xe & 200 & $4\times10^{25}$ & 100--270 & current \\
& & 5000 & $10^{27}-10^{28}$ & 6--53 & R\&D \\ 
SuperNEMO & $^{82}$Se & 7 & 6.5$\times10^{24}$ & 240--570 & in progress \\
 \cite{GOM16} &  & 100--200 & (1--2)$\times10^{26}$ & 40--140 & R\&D \\
KamLAND-Zen  & $^{136}$Xe & 750 & 2$\times10^{26}$ & 45--120 & in progress \\
 \cite{GAN16a}& & 1000 &  $6\times10^{26}$ & 26-69 & R\&D \\
SNO+ \cite{AND16} & $^{130}$Te & 800 & $9\times10^{25}$ & 55--205 & in progress \\
 & & 8000 & $7\times10^{26}$ & 20-73 &  R\&D \\
\hline
\end{tabular}
%\end{center}
\end{table}

\subsection{Near future (2017-2019)}
I believe that to the end of this period we will have the following results:\\
1. $^{}$Xe: $T_{1/2}$(0$\nu$) $> 2\cdot10^{26}$ yr ($\langle m_{\nu} \rangle$ $< 0.045-0.120$ eV).

This result will be obtained by new KamLAND-Zen experiment with 750 kg of $^{136}$Xe and with new (more pure) internal balloon.\\
2. $^{76}$Ge: $T_{1/2}$(0$\nu$) $> 1.5\cdot10^{26}$ yr ($\langle m_{\nu} \rangle$ $< 0.09-0.23$ eV). 

This result will be obtained combining GERDA-II ($\sim 10^{26}$ yr) and MAJORANA-DEMONSTRATOR ($\sim 10^{26}$ yr) results.\\ 
3. $^{130}$Te: $T_{1/2}$(0$\nu$) $> 1\cdot10^{26}$ yr ($\langle m_{\nu} \rangle$ $< 0.05-0.19$ eV).

This result will be obtained combining CUORE ($\sim 0.7\cdot10^{26}$ yr) and SNO+ ($\sim 0.7\cdot10^{26}$ yr) results. 

Of course, during this period some other new results will be obtained (EXO, SuperNEMO-Demonstrator, NEXT, LUCIFER, LUCINEU, AMORE,...), but it will be not competitive with mentioned above results (in the sense of sensitivity to $\langle m_{\nu} \rangle$).

\subsection{Far future (2020-2030)}
It is quite complicated to predict far future. This is why I will try just to predict dates when the most prospect experiments will start to take data. One can find these predictions in table 5.
\begin{table}[ht]
\label{Table1}
\caption{Start of data taking for some large scale experiments (prediction).}
\vspace{0.5cm}
%\rule[-2mm]{0mm}{5mm}
\begin{center}
\begin{tabular}{cc}
\hline
Experiment & Start of data taking, yr \\
\hline
KamLAND2-Zen (1000 kg of $^{136}$Xe) & $\sim 2020-2022$ \\
SNO+ (8000 kg of $^{nat}$Te) & $\sim 2020-2022$ \\
CUPID ($^{100}$Mo, $^{82}$Se, $^{116}$Cd,...) & $\sim 2022$ \\
LEGEND-I (200 kg of $^{76}$Ge) & $\sim 2022-2025$ \\
LEGEND (1000 kg of $^{76}$Ge) & $\sim 2025-2030$ \\
nEXO (5000 kg of $^{136}$Xe) & $\sim 2025-2030$ \\
\hline
\end{tabular}
\end{center}
\end{table}
\section{Conclusion}
Thus, at present, the $2\nu\beta\beta$ decay of 11 nuclei ($^{48}$Ca, $^{76}$Ge, $^{82}$Se,
$^{96}$Zr, $^{100}$Mo, $^{116}$Cd, $^{128}$Te, $^{130}$Te, $^{136}$Xe, $^{150}$Nd and $^{238}$U) 
has been registered. Moreover, the $2\nu\beta\beta$
decay of $^{100}$Mo and $^{150}$Nd to 0$^+_1$ excited states of daughter nuclei and the ECEC(2$\nu$)
process in $^{130}$Ba (geochemical experiments) have been detected too.

The $0\nu\beta\beta$ decay has not been observed yet, and the best limits on $\langle m_{\nu} \rangle$
have been obtained in experiments with $^{136}$Xe, $^{76}$Ge, $^{100}$Mo, and $^{130}$Te.
Using most reliable NME calculations it is possible to set the present conservative limit as 
$\langle m_{\nu} \rangle < 0.24$ eV. Conservative present limit on decay with Majoron 
emission has been obtained as $\langle g_{ee} \rangle < 1.2\cdot10^{-5}$ (ordinary Majoron with n = 1).

Sensitivity to $\langle m_{\nu} \rangle$ on the level $\sim$ 0.05-0.20 eV will 
be reached by next generation experiments 
in a few years from now and on the level $\sim$ 0.02-0.05 eV (inverted hierarchy region) after 2022. I do not discuss here effect of possible quenching of axial-vector coupling constant $g_A$ (see discussions in \cite{BAR15,PIR15}, for example). If the quenching of $g_A$ really exist it will decrease our sensitivity to $\langle m_{\nu} \rangle$.

\ack
%\subsection{Acknowledgments}

Portions of this work were supported by 
grants from RFBR (no 15-02-0291915).

\end{document}